This paper is appearing in the *Finance Research Letters*. Please cite this article as: Ormos, M., Timotity, D., Market Microstructure During Financial Crisis: Dynamics of Informed and Heuristic-Driven Trading, Finance Research Letters (2016), DOI: 10.1016/j.frl.2016.06.003
This is the pre-print version of our accepted paper before typesetting.

# Market Microstructure During Financial Crisis
## Dynamics of Informed and Heuristic-Driven Trading[1]

**MIHÁLY ORMOS**
Department of Finance, Budapest University of Technology and Economics
Magyar tudosok krt. 2., 1117 Budapest, Hungary
Phone: +36 1 4634220; Fax: +36 1 4632745
ormos@finance.bme.hu

**DUSAN TIMOTITY**
Department of Finance, Budapest University of Technology and Economics
Magyar tudosok krt. 2., 1117 Budapest, Hungary
Phone: +36 1 4634220; Fax: +36 1 4631083
timotity@finance.bme.hu

## Abstract

We implement a market microstructure model including informed, uninformed and heuristic-driven investors, which latter behave in line with loss-aversion and mental accounting. We show that the probability of informed trading (PIN) varies significantly during 2008. In contrast, the probability of heuristic-driven trading (PH) remains constant both before and after the collapse of Lehman Brothers. Cross-sectional analysis yields that, unlike PIN, PH is not sensitive to size and volume effects. We show that heuristic-driven traders are universally present in all market segments and their presence is constant over time. Furthermore, we find that heuristic-driven investors and informed traders are disjoint sets.

---

[1] **Acknowledgements:** We highly appriciate the comments and suggestions of Niklas Wagner. Mihály Ormos acknowledges the support by the János Bolyai Research Scholarship of the Hungarian Academy of Sciences. Dusán Timotity acknowledges the support by the Pallas Áthéné Domus Scientiae Foundation. This research was partially supported by Pallas Áthéné Domus Scientiae Foundation.

**Introduction**

Market microstructure has long been considered to influence asset pricing (Glosten and Milgrom 1985; Kyle, 1985). One of the key points in this relationship lies in asymmetric allocation of information in capital markets contributing to the uncertainty around the investments of liquidity traders. Easley et al. (1996, 2002) capture this latter effect by introducing a novel measure, the probability of informed trading (PIN), and show its significant effect on the liquidity and the expected return of assets.

In this paper, we describe the dynamics of their proposed measure within a sample including highly volatile periods and structural shocks: we analyze the microstructural change during 2008 and take into consideration the structures of both the pre- and post-Lehman era. Our sample covers every executed trade made through Budapest Stock Exchange between 2 January 2008 and 31 December 2008.

Furthermore, we extend our analysis to another type of investors. According to Ormos and Timotity (2016a), due to loss-aversion and intertemporal mental accounting, investors tend to follow contrarian strategies: they invest into riskier (less risky) assets or increase (decrease) leverage subsequent to previous negative (positive) market shocks. We introduce this latter pattern into market microstructure by including a class of heuristic-driven traders in the base model consisting of informed and uninformed traders and specialists. The relevance of such traders in market microstructural analysis is supported by numerous studies: according to Yao and Li (2013), prospect theory investors can behave as contrarian noise traders in a market. This is confirmed by Chordia et al. (2002), who find that order imbalance increases (decreases) subsequent to market declines (jumps), which indicates contrarian investors in aggregate. Furthermore, providing evidence for our particular explanation, Kadous et al. (2014) argue that investors act as contrarians only if they have held the particular asset in the past that they buy in the subsequent period.

In our empirical tests we find that, while the proportion of informed investors varies in time and amongst different assets, the probability of heuristic-driven trading (PH) is insensitive to temporal and cross-sectional factors. Therefore, we argue that the existence of heuristic-driven traders forms a universal and robust property of capital markets.

## 1. Data and Methodology

In contrast to Glosten and Milgrom (1985) and Kyle (1985), where uninformed traders are defined as those who do not possess fundamental information on assets, irrespective of their motives, our setting is rather similar to the paper of Bloomfield et al. (2009), in which uninformed investors can have other trading motives than fundamental (e.g. behavioral). In particular, we argue that there exists a class of heuristic-driven trader that behaves in line with the contrarian investment pattern (i.e. buy/sells subsequent to losses/gains), which leads to a positive probability of such investors. We apply this probability (PH) as a defining measure of market microstructure.

This inclusion of PH in our microstructure model is based on the following idea by Ormos and Timotity (2016a): investors that are sensitive to loss-aversion and mental accounting, when they lose (gain) money, tend to aggregate in time, and therefore, their required return increases (decreases) in the subsequent period; then, this raised (lowered) expected return can be obtained by investing in riskier assets or increasing leverage, hence, their demand for risky assets increases (decreases). Altogether, therefore, this behavior leads to a negative relationship between order imbalance and previous market returns.

This pattern could also provide an explanation for recent findings on investors' portfolio choice, according to which high-aspiration investors (those with high required returns) trade more than the average (Magron, 2014). For example, if one invests in assets providing higher expected return with greater risk, unexpected deviations become larger; hence, the aforementioned pattern plays a more significant role in portfolio choice.

Our microstructural model is based on the following assumptions: (1) investors hold well-diversified portfolios; therefore, changes of the market portfolio is a proxy for their portfolio; (2) all stocks are risky assets; thus, subsequent to a market loss (gain), investors increase (decrease) their portfolio risk and expected return through leverage in the market portfolio, which yields a higher (lower) demand for individual stocks; (3) orders from heuristic-driven buyers have Poisson distribution and arrive at rate $\epsilon_b^H \max(0, -I_{t-1})$, while order from heuristic-driven sellers arrive at rate $\epsilon_s^H \max(0, I_{t-1})$, where $I_{t-1}$ stands for an indicator function of

$$I_{t-1} = \begin{cases} 1 \text{ if } r_{t-1}^M \geq 0 \\ -1 \text{ if } r_{t-1}^M < 0 \end{cases},$$

in which $r_{t-1}^M$ stands for the previous market return. Since we use the daily number of buy and sell transactions in PIN estimations, the length of the previous market shock is also set to one day (thus, $I_{t-1}$ is the function of the return of the preceding day). This is in line with the empirical findings on the inititally strong, yet fading property of anchoring to previous returns (Ormos and Timotity, 2016b). The intuition behind this latter assumption is defined by the contrarian pattern discussed above, which aims to capture the increased number of buy (sell) orders subsequent to market losses (gains).

This asymmetric framework is in line with recent studies on the relationship between the market return and future trading volume: according to Dodonova (2015), extreme negative market returns lead to high future trading volume, while extreme positive returns have only a slight effect on future volume. Our second assuption is also supported by an extensive amount of results on the hedging role of government bonds in portfolio diversification, which is especially relevant in highly volatile periods (Acosta-González et al., 2015).

Apart from the aforementioned, we apply a set of parameters similar to Easley et al. (1996, 2002): private information emerges each day with probability α, which contains bad and good news with probabilities δ and (1-δ) respectively. Orders from uninformed buyers and sellers arrive at rate $\epsilon_b$ and $\epsilon_s$, while informed orders arrive at rate μ if the private information event has occurred. B and S represent total buy trades and sell trades for the day respectively. Then, the likelihood function of the parameter set θ for a single day is given by:

$$L(\theta|B,S,I_{t-1}) = (1-\alpha)e^{-[\epsilon_b+\epsilon_b^H \max(0,-I_{t-1})]}\frac{[\epsilon_b+\epsilon_b^H \max(0,-I_{t-1})]^B}{B!}e^{-[\epsilon_s+\epsilon_s^H \max(0,I_{t-1})]}\frac{[\epsilon_s+\epsilon_s^H \max(0,I_{t-1})]^S}{S!} +$$
$$+\alpha\delta e^{-[\epsilon_b+\epsilon_b^H \max(0,-I_{t-1})]}\frac{[\epsilon_b+\epsilon_b^H \max(0,-I_{t-1})]^B}{B!}e^{-[\mu+\epsilon_s+\epsilon_s^H \max(0,I_{t-1})]}\frac{[\mu+\epsilon_s+\epsilon_s^H \max(0,I_{t-1})]^S}{S!} +$$
$$+\alpha(1-\delta)e^{-[\mu+\epsilon_b+\epsilon_b^H \max(0,-I_{t-1})]}\frac{[\mu+\epsilon_b+\epsilon_b^H \max(0,-I_{t-1})]^B}{B!}e^{-[\epsilon_s+\epsilon_s^H \max(0,I_{t-1})]}\frac{[\epsilon_s+\epsilon_s^H \max(0,I_{t-1})]^S}{S!}. \quad (1)$$

Assuming independence between the number of daily buy and sell orders yields the likelihood function for n days:

$$L(\theta|D) = \prod_{n=1}^{N} L(\theta|B_n, S_n, r_{n-1}^M), \quad (2)$$

where D stands for the dataset consisting of the B, S and $r^M$ series for the n=1,..,N interval. We use the average of the daily values for PIN calculations such that

$$PIN = \frac{1}{N}\sum_{n=1}^{N}\frac{\alpha\mu}{\alpha\mu+\epsilon_b+\epsilon_s+\epsilon_b^H \max(0,-r_{t-1}^M)+\epsilon_s^H \max(0,-r_{t-1}^M)}, \quad (3)$$

whereas PH is defined as

$$PH = \frac{1}{N}\sum_{n=1}^{N}\frac{\epsilon_b^H \max(0,-r_{t-1}^M)+\epsilon_s^H \max(0,-r_{t-1}^M)}{\alpha\mu+\epsilon_b+\epsilon_s+\epsilon_b^H \max(0,-r_{t-1}^M)+\epsilon_s^H \max(0,-r_{t-1}^M)}. \quad (4)$$

According to Yan and Zhang (2012), annual periods are too long to capture the microstructure, and may not reflect the dynamics chaning at quarterly earnings reports; as the Hungarian reporting system is similar to the one in the US in this sense, following their methodology we apply quarterly periods. This method also allows us to run a time-series analysis that captures the dynamics of market microstructure.

Maximum likelihood estimation of the parameters is highly dependent on the starting point of such complex structures; in order to avoid finding a local maximum, we use Monte Carlo simulation. For each period and asset 10,000 parameter sets are simulated with uniform distributions under the following constraints: $\epsilon_b, \epsilon_b^H \in [0, \bar{B}]; \epsilon_s, \epsilon_s^H \in [0, \bar{S}]; \alpha, \delta \in [0,1]; \mu \in \left[\frac{\bar{S}}{\alpha\delta}\right]$.

Data covers the 24,964,980 trades made through the Budapest Stock Exchange between 2 January 2008 and 31 December 2008 on 644 assets. In the PIN and PH estimations, we filter out non-equity assets and those not having at least one sell and buy transaction in each trading day; the remaining sample consists of 11,762,646 trades on 45 equities. The specific statistics of these equities are shown in Appendix 1.

2. **Empirical Results**

Table 1 shows the results of parameter estimations. The descriptive statistics of commonly used variables, such as the PIN, indicate similar findings to existing literature (see Easley et al., 2002). The novelty in our estimation, the PH variable, however, indicates a significant proportion of heuristic-driven trading activity in capital markets. Based on the estimated probabilities, for almost every third private information-based trade, there is one heuristic-driven transaction.

**Please Insert Table 1 here**

We present our results on the dynamics of PIN in Table 2. We argue that subsequent to the Lehman collapse at the end of Q3, when shocks and high volatility were persistent; the proportion of informed traders in the market decreased significantly. This finding is somewhat surprising considering that recent results have revealed a higher effect of information asymmetry during recession (Lin, 2015), which was particulary strong in the Q4 of 2008.

**Please Insert Table 2 about here**

Table 3 shows that change in PIN cannot be attributed to a change in the probability of heuristic-driven trades but to the third class, the uninformed liquidity traders. In contrast to PIN estimations, PH in the fourth quarter, surprisingly, does not change significantly. In fact, we find that PH is insensitive to even huge structural shocks, and is constant in time.

**Please Insert Table 3 about here**

In the followings we present the robustness of our results in a cross-sectional analysis. In line with Easley et al. (2002) we test the size effect, which modifies market microstructure. The importance of this analysis is highlighted by recent results on the size effects on asset prices. First, Ortiz et al. (2015) find no return anomalies in the first three quarters of their sample years; however, in the fourth quarter they find a significant turn-of-the-year effect for small-cap stocks with poor performance, which could influence our results as well. Second, Lou and Schinckus (2015) document an asymmetric effect of small and large stocks indicating that herding behavior is particularly present for small stocks in bullish periods, while in bearish times crowd moves for large shares, which in our analysis might lead to biased estimations of the microstructure without controlling for the size effect. We represent our findings in Figure 1, which shows the PIN and PH estimations of nine groups in increasing order based on market capitalization. Each group contains five assets based on its ranking (i.e. group 1 stands for the smallest five firms). One can clearly see a decreasing linear fit of PIN on the size of the group, whereas, a horizontal line is fitted for the PH indicating no significant relationship. The exact numbers for each firm are shown in Appendix 1.

**Please Insert Figure 1 about here**

In particular, as the p-value indicates in Table 4, we indeed find the usual, significant negative relationship between the market capitalization and the probability of informed trading related to a given asset. Still, the significance vanishes if we change the dependent variable to PH.

**Please Insert Table 4 about here**

Furthermore, in order to increase the robustness of results, we test the cross-sectional difference by applying mean daily volume instead of market capitalization. Nevertheless, we find similar patterns: as presented in Table 5, PH is not significant at reasonable probability levels; however, the relationship is negative between PIN and volume and is very significant.

**Please Insert Table 5 about here**

Finally, we also investigate the question whether the classes of informed and heuristic traders are disjoint sets and independent of each other. In order to do so, we add the PIN values to the PH estimation in addition to the latter two control variables for cross-sectional differences, and we also introduce a dummy variable for the fourth quarter to account for temporal differences in the post-Lehman recession.

Our results shown in Table 6 indicate that time-series and cross-sectional control variables have no influence on the probability of heuristic driven trading; however, it is negatively and significantly affected by the probability of informed trading. This finding confirms that informed and heuristic traders are two disjoint sets, and have no common elements, which would otherwise yield a positive relationship between the two probabilities. Furthermore, we attribute no economical relevance to this negative relationship, since the magnitude of the effect indicates only a marginal (-0.05 percent) change of PH for a one percent change in PIN. Therefore, we argue that heuristic driven trading activity is independent of time-series and cross-sectional effects and information asymmetry.

**Please Insert Table 6 about here**

According to Kaul et al. (2008), estimations of relationship between the probability of informed trading and volume and the PIN itself are sensitive to the sampling period applied. Therefore, we also test the robustness of our previous results using monthly sampling periods, which, although yields a lower number of observations in the MLE, allows for better capturing the time-series dynamics of market microstructure.

Nevertheless, as shown in Table 7, we still obtain an almost exact match to the quarterly results with the statistics related to PH and PIN estimations; therefore, we argue that the aforementioned findings are also robust with respect to the sampling horizon. Moreover, regression estimations for monthly horizons yield that volume loses its significance even in the PIN estimation, whereas, the PH estimation indicates no significant effect of PIN, size, volume, or the dummy variable for September 2008. These results further support our findings on the disjoint sets of heuristic traders and information-based investors.

**Please Insert Table 7 about here**

3. **Concluding Remarks**

We investigate the dynamics of market microstructure during 2008. The analysis of this period allows measuring the effect of great structural shocks, such as the one beginning with the collapse of Lehman Brothers in September 2008. Our analysis yields four novel findings.

First, we measure a significant decrease in the probability of informed trading subsequent to the global plunge in stock prices in September. According to our results, this can be attributed to an increase in uninformed liquidity trading but not to heuristic-driven (or contrarian) trading. Thus, in periods of great uncertainty, when prices are highly volatile, rational, value-focused investors tend to stay away from the market; even though, large deviations from fundamental prices offer investment opportunities with high expected return.

Second, in contrast to PIN estimations, we find no significant temporal difference between the probabilities of heuristic-driven trading, which latter is reflected by contrarian behavior. In other words, PH is insensitive to widespread market shocks and to changes in volatility; hence, we argue that this measure is robustly constant over time.

Third, in cross-sectional analysis, we find that PIN is affected negatively by both the market capitalization and average volume of assets. For PH, however, no significant cross-sectional effect has been found using these two variables, even if controlling for information asymmetry as well. Therefore, in addition to temporal robustness, PH is constant cross-sectionally as well; morever, this latter class of investors forms a set independent of informed traders as the two are disjoint.

Finally, both the quarterly and monthly sampled analysis confirms that PIN and PH are not related to each other with economic significance; that is, heuristic-driven or contrarian traders form an independent and disjoint set from information-based investors.

Altogether, we argue that heuristic-driven trading forms a universal and robust property of capital markets even in highly volatile periods and during significant changes, whereas, informed trading has robust dynamics in both cross-sectional and time-series analysis.

# Tables

**Table 1: Parameter estimation results**

|  | $\epsilon_b$ | $\epsilon_s$ | $\epsilon_b^H$ | $\epsilon_s^H$ | $\alpha$ | $\delta$ | $\mu$ | PH | PIN |
|---|---|---|---|---|---|---|---|---|---|
| Mean | 385.25 | 500.29 | 53.44 | 56.64 | 0.42 | 0.56 | 288.76 | 0.06 | 0.19 |
| Median | 35.87 | 68.30 | 8.61 | 6.54 | 0.41 | 0.57 | 84.95 | 0.05 | 0.16 |
| Std. Dev. | 1552.81 | 1678.06 | 152.52 | 215.53 | 0.23 | 0.25 | 858.23 | 0.03 | 0.11 |
| 10$^{th}$ percentile | 6.02 | 5.22 | 0.43 | 0.38 | 0.14 | 0.22 | 10.03 | 0.02 | 0.07 |
| 90$^{th}$ percentile | 573.28 | 804.25 | 138.62 | 93.44 | 0.77 | 0.91 | 616.69 | 0.10 | 0.34 |

*Notes: The numbers represent the statistics of MLE optimized parameters for the whole period. The parameters are estimated on quarterly intervals.*

**Table 2: Differences in quarterly means of PIN in 2008**

|  | Q1 | Q2 | Q3 | Q4 |
|---|---|---|---|---|
| Q1 | 0 | 0.0289 | -0.0029 | -0.0500* |
| Q2 |  | 0 | -0.0318 | -0.0788** |
| Q3 |  |  | 0 | -0.0471* |
| Q4 |  |  |  | 0 |

*Notes: The values show the differences between average PINs in the quarter indicated in the top row versus the quarter indicated in the left column. * and ** stand for significant difference at 5% and 1% respectively.*

**Table 3: Differences in quarterly means of PH**

|  | Q1 | Q2 | Q3 | Q4 |
|---|---|---|---|---|
| Q1 | 0 | 0.0021 | 0.0052 | 0.0093 |
| Q2 |  | 0 | 0.0030 | 0.0072 |
| Q3 |  |  | 0 | 0.0042 |
| Q4 |  |  |  | 0 |

*Notes: The numbers represent the quarterly differences of PH throughout 2008. Due to symmetry, the lower triangular values are omitted.*

**Table 4: Effect of market capitalization on the microstructure**

|  | PH | | PIN | |
|---|---|---|---|---|
|  | Coefficient | P-value | Coefficient | P-value |
| Constant | 0.0565 | 0.0000 | 0.1970 | 0.0087 |
| Mkt.cap | 5.41E-16 | 0.8939 | -5.21E-14 | 0.0008 |

*Notes: The regressions applied are $PH_i = \alpha + \beta \cdot mktcap_i + \varepsilon_i$ and $PIN_i = \alpha + \beta \cdot mktcap_i + \varepsilon_i$. P-values stand for the probability of rejecting a true null hypothesis using Student's t-distribution.*

**Table 5: Effect of trading volume on the microstructure**

|  | PH | | PIN | |
|---|---|---|---|---|
|  | Coefficient | P-value | Coefficient | P-value |
| Constant | 0.0566 | 0.0000 | 0.1958 | 0.0000 |
| Volume | -2.36E-10 | 0.9357 | -3.17E-08 | 0.0048 |

*Notes: The regressions applied are $PH_i = \alpha + \beta \cdot volume_i + \varepsilon_i$ and $PIN_i = \alpha + \beta \cdot volume_i + \varepsilon_i$. P-values stand for the probability of rejecting a true null hypothesis using Student's t-distribution.*

**Table 6: Effect of trading volume on the microstructure**

|  | PH | |
|---|---|---|
|  | Coefficient | P-value |
| Constant | 0.0667 | 0.0000 |
| Mkt.cap | 8.75E-16 | 0.9171 |
| Volume | -2.24E-09 | 0.7090 |
| Q4 | -0.0042 | 0.4111 |
| PIN | -0.0459 | 0.0250 |

*Notes: The regression applied is $PH_{i,t} = \alpha + \beta_1 \cdot mkt.cap_{i,t} + \beta_2 \cdot volume_{i,t} + \beta_3 \cdot D_{Q4,i,t} + \beta_4 \cdot PIN_{i,t} + \varepsilon_i$. P-values stand for the probability of rejecting a true null hypothesis using Student's t-distribution.*

**Table 7: Parameter estimation results with monthly sampling periods**

|  | $\epsilon_b$ | $\epsilon_s$ | $\epsilon_b^H$ | $\epsilon_s^H$ | $\alpha$ | $\delta$ | $\mu$ | PH | PIN |
|---|---|---|---|---|---|---|---|---|---|
| Mean | 377.14 | 467.17 | 96.11 | 95.49 | 0.52 | 0.56 | 288.76 | 0.06 | 0.19 |
| Median | 32.5 | 61.64 | 9.33 | 11.5 | 0.54 | 0.61 | 75.27 | 0.08 | 0.19 |
| Std. Dev. | 1559.39 | 1587.03 | 152.52 | 283.78 | 0.25 | 0.25 | 4730.46 | 0.05 | 0.13 |
| 10th percentile | 5.83 | 5.09 | 0.76 | 0.79 | 0.17 | 0.24 | 6.87 | 0.03 | 0.06 |
| 90th percentile | 475.5 | 803.08 | 199.51 | 226.3 | 0.85 | 0.91 | 646.35 | 0.17 | 0.36 |

*Notes: The numbers represent the statistics of MLE optimized parameters for the whole period. The parameters are estimated on monthly intervals.*

# Figures

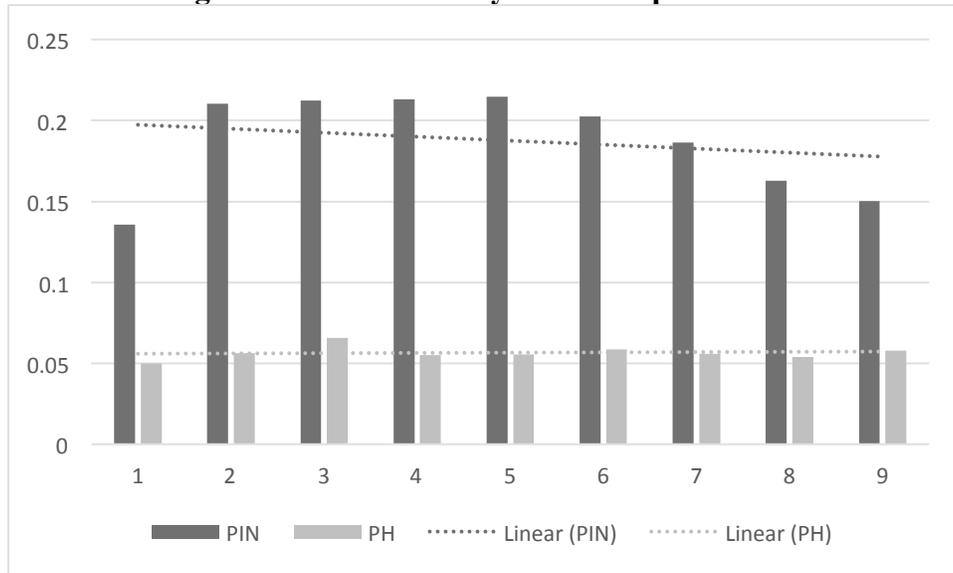

**Figure 1: PIN and PH by market capitalization**

*Notes: PIN and PH estimations are represented for nine groups, which latter contain five assets each, based on their market capitalization in increasing order.*

# Appendix 1

**Table A/1: PIN and PH estimations and cross-sectional statistics for sample firms**

| Ticker | PIN | PH | Market capitalization (in million HUF) | Total number of transactions in the sample |
|---|---|---|---|---|
| FEVITAN | 0.0781 | 0.0316 | 745 | 2753 |
| EHEP | 0.0538 | 0.0700 | 888 | 2434 |
| CSEPEL | 0.1439 | 0.0390 | 1087 | 13196 |
| KONZUM | 0.2587 | 0.0463 | 1502 | 31959 |
| ERSTEALAPG05 | 0.1447 | 0.0620 | 2766 | 2273 |
| PFLAX | 0.1553 | 0.0530 | 2791 | 34034 |
| OTPALFA2 | 0.2241 | 0.0459 | 2838 | 5220 |
| OTPLINEA3 | 0.2130 | 0.0771 | 3290 | 3933 |
| HUMET | 0.1671 | 0.0561 | 3375 | 262163 |
| BOOK | 0.2917 | 0.0497 | 4176 | 26373 |
| PHYLAXIA | 0.2168 | 0.0453 | 4422 | 262323 |
| ETFBUXOTP | 0.1922 | 0.1093 | 4739 | 362097 |
| TVNETWORK | 0.2634 | 0.0474 | 4819 | 16651 |
| RFV | 0.1963 | 0.0749 | 5000 | 21704 |
| FREESOFT | 0.1930 | 0.0520 | 5259 | 31084 |
| ERSTEALAPG04 | 0.1546 | 0.0472 | 5649 | 3608 |
| QUAESTOR | 0.2059 | 0.0619 | 6338 | 5687 |
| FORRAS/OE | 0.1856 | 0.0622 | 6812 | 30620 |
| OTPLINEA | 0.2445 | 0.0549 | 6937 | 5954 |
| ECONET | 0.2744 | 0.0492 | 7632 | 220195 |
| MKBPAGODA | 0.1570 | 0.0649 | 11530 | 5230 |
| FORRAS/T | 0.1865 | 0.0603 | 12395 | 19955 |
| GSPARK | 0.1561 | 0.0745 | 14087 | 31727 |
| SYNERGON | 0.2825 | 0.0302 | 17084 | 181501 |
| GENESIS | 0.2913 | 0.0479 | 17668 | 171423 |
| OTPALFA3 | 0.1794 | 0.0589 | 17759 | 13123 |
| OTPOSZTCS1 | 0.1586 | 0.0681 | 17876 | 9133 |
| ANY | 0.2072 | 0.0582 | 19761 | 65979 |
| RABA | 0.3127 | 0.0557 | 25330 | 244512 |
| AAA | 0.1538 | 0.0528 | 28458 | 68484 |
| BIF | 0.1530 | 0.0613 | 29681 | 9967 |
| LINAMAR | 0.2139 | 0.0663 | 30888 | 15365 |
| ZWACK | 0.2234 | 0.0367 | 32320 | 28237 |
| FOTEX | 0.1450 | 0.0690 | 61530 | 236716 |
| DANUBIUS | 0.1967 | 0.0464 | 74569 | 50186 |
| EMASZ | 0.2054 | 0.0580 | 76261 | 57569 |
| FHB | 0.1584 | 0.0533 | 105000 | 206922 |
| EGIS | 0.1823 | 0.0335 | 145000 | 328751 |
| TVK | 0.1331 | 0.0562 | 171000 | 105823 |
| ELMU | 0.1345 | 0.0693 | 187000 | 27658 |
| ORC | 0.3323 | 0.0585 | 223000 | 259583 |
| RICHTER | 0.1216 | 0.0647 | 757000 | 775659 |
| MTELEKOM | 0.1703 | 0.0512 | 918000 | 1026163 |
| OTP | 0.0542 | 0.0523 | 2480000 | 4792047 |
| MOL | 0.0724 | 0.0633 | 2610000 | 1686672 |